\documentclass[12pt]{article}
\usepackage{amssymb}

\usepackage{amsmath,amssymb,amsbsy,amsfonts,amsthm,latexsym,amsopn,amstext,amsxtra,euscript,amscd,color}
\usepackage{color,xcolor}
\usepackage{amsfonts}
\usepackage{amsfonts,mathrsfs}
\usepackage{psfrag}
\usepackage{subfigure}
\usepackage{url}
\usepackage{stfloats}
\usepackage{amsmath}
\usepackage{amssymb}

\usepackage{latexsym}

\newtheorem{theorem}{Theorem}
\newtheorem{lemma}{Lemma}

\begin{document}

\title{On the $k$-error linear complexity of binary sequences derived from the discrete logarithm in finite fields}

\author{Zhixiong Chen$^{1}$ and Qiuyan Wang$^{1,2}$
\\
1. Provincial Key Laboratory of Applied Mathematics,
\\ Putian University,
\\ Putian, Fujian 351100, P.R. China\\
ptczx@126.com\\
2. School of Computer Science and Software Engineering,\\
 Tianjin Polytechnic University,\\
 Tianjin 300387, P.R. China
}

\maketitle
\begin{abstract}
Let $q=p^r$ be a power of an odd prime $p$.
We study binary sequences $\sigma=(\sigma_0,\sigma_1,\ldots)$ with entries in $\{0,1\}$ defined by using the
 quadratic character $\chi$ of the finite field $\mathbb{F}_q$:
$$
\sigma_n=\left\{
\begin{array}{ll}
0,& \mathrm{if}\quad n= 0,\\
(1-\chi(\xi_n))/2,&\mathrm{if}\quad 1\leq n< q,
\end{array}
\right.
$$
for the ordered elements $\xi_0,\xi_1,\ldots,\xi_{q-1}\in \mathbb{F}_q$. The  $\sigma$ is Legendre sequence if $r=1$.

 Our first contribution is to prove a lower bound on the linear complexity of $\sigma$ for $r\geq 2$.
 The bound improves some results of Meidl and Winterhof. Our second contribution is to study the $k$-error linear complexity of $\sigma$ for $r=2$.
It seems that we cannot settle the case when $r>2$ and leave it open.  \\

\textbf{keyword:} stream cipher;  pseudorandom binary sequences; linear complexity;  $k$-error linear complexity; discrete
logarithm; finite field
\end{abstract}

\section{Introduction}

Pseudorandom sequences play an important role in cryptography. In particular in  symmetric cryptography they serve as the secret key.
So pseudorandom sequences are widely concerned. In this work, we begin with the Legendre sequence which has good behavior.

Let $p$ be an odd prime. The \emph{Legendre sequence} $\ell=\{\ell_0,\ell_1,\ldots\}$ with entries in  $\{0,1\}$ is defined as
$$
\ell_n=\left\{
\begin{array}{ll}
0, & \mathrm{if}\,\ n\equiv 0\pmod p,\\
\frac{1-\left(\frac{n}{p}\right)}{2}, & \mathrm{otherwise},
\end{array}
\right.  ~~ n \ge 0,
$$
where $\left(\frac{\cdot}{p}\right)$ is the Legendre symbol, that is for $n$ with $p\nmid n$, $\left(\frac{n}{p}\right)=1$ if $n\equiv a^2 \pmod p$ for some integer $a$ and otherwise $\left(\frac{n}{p}\right)=-1$.
The Legendre sequence is extensively paid attention by many researchers. From the viewpoint of cryptography,
the {\em linear complexity} (see the notion below) of it is studied in \cite{DHS}, the \emph{$k$-error linear complexity} of it is studied in \cite{AW}\footnote{We remark that the Legendre sequence is treated as a $p$-ary sequence over $\mathbb{F}_p$.},
and other feathers are studied  in the literature, see e.g., \cite{CDR,KS}.

It is natural to extend the Legendre symbol construction to define binary sequences from the extension field $\mathbb{F}_q$ of $q$ elements
with $q=p^r$. We order  the elements of $\mathbb{F}_q=\{\xi_0,\xi_1,\ldots,\xi_{q-1}\}$ as follows.

Fixing a basis $\{\gamma_1=1, \gamma_2,\ldots,\gamma_r\}$  of $\mathbb{F}_q$ over
$\mathbb{F}_p=\{0,1,\ldots,p-1\}$, we define for $0\leq n<q$,
$$
\xi_n=n_1\gamma_1+n_2\gamma_2+\cdots+n_r\gamma_r,
$$
if
$$
n=n_1+n_2p+\cdots+n_rp^{r-1}, \ 0\leq n_i<p, \quad i=1,\ldots,r.
$$

Let $\alpha$ be a primitive element of $\mathbb{F}_q$ and $\mathrm{ind}_{\alpha}(\xi)\in \mathbb{Z}_{q-1}=\{0,1,\ldots,q-2\}$ the
\emph{discrete logarithm} of $0\neq \xi\in \mathbb{F}_q$ with respect to $\alpha$, i.e.,
$\xi=\alpha^{\mathrm{ind}_{\alpha}(\xi)}$. For any integer $d>1$, we use the notation
$\mathrm{ind}_{\alpha,d}(\xi):=\mathrm{ind}_{\alpha}(\xi) \pmod d$.
 The importance of the discrete logarithm for modern cryptography is well
known. The security of many public-key cryptosystems depends on the intractability of the discrete logarithm problem.

The $q$-periodic sequence $\sigma=\{\sigma_0,\sigma_1,\ldots\}$ with entries in  $\mathbb{Z}_d=\{0,1,\ldots,d-1\}$ defined below has been concerned in the literature:
\begin{equation}\label{ind}
\sigma_n=\left\{
\begin{array}{ll}
0,& \mathrm{if}\quad n=0,\\
\mathrm{ind}_{\alpha,d}(\xi_n),&\mathrm{if}\quad 1\leq n< q,
\end{array}
\right.\mathrm{and} \quad \sigma_{n+q}=\sigma_{n}, n\geq 0.
\end{equation}
It is clear $\sigma$ is the Legendre sequence if $r=1$ and $d=2$. The (aperiodic) autocorrelation of $\sigma$ was
analyzed in \cite{MW032} and the  linear complexity  of $\sigma$ was
studied in \cite{MW01,W04}.
 In particular,  in \cite{AMW,AW} the $k$-error linear
complexity over $\mathbb{F}_p$ of $\sigma$ was investigated for $r=1$.
One might ask whether it can be extended to the case $r\geq 2$ for the $k$-error linear complexity.
Indeed, \cite[Proposition 2]{BEP} tells us that we have
to change a lot of elements to get a smaller periodic sequence with small linear complexity.
So this might be the reason why  the authors of \cite{AMW,AW}
not study $\sigma$ over $\mathbb{F}_p$  furtherly for the case $r\geq 2$.

In this work, we pay attention to  the case when $d=2$.
The $\sigma$ with entries in $\{0,1\}$ in Eq.(\ref{ind}) can be defined equivalently by using the \emph{quadratic character} $\chi$ of $\mathbb{F}_q$:
\begin{equation}\label{chi}
\sigma_n=\left\{
\begin{array}{ll}
0,& \mathrm{if}\quad n= 0,\\
(1-\chi(\xi_n))/2,&\mathrm{if}\quad 1\leq n< q.
\end{array}
\right.
\end{equation}
It is easy to see that $\chi(\xi_n)=(-1)^{\sigma_n}$ for $1\leq n<q$. The measures of pseudorandomness of the binary $\sigma$ was studied in
\cite{SW} for more general setting. Some related problems were considered in \cite{GMS,GSS09,GSS13}.
 In the sequel, we first prove a lower bound on the linear complexity of of $\sigma$ in Eq.(\ref{chi}) for $q=p^r$ with $r\geq 2$ in Sect.2.
 The bound improves some results of Meidl and Winterhof in \cite{MW01,W04}.
 Then  in Sect.3 we  study its $k$-error linear complexity (over $\mathbb{F}_2$) of $\sigma$ for $r=2$.
This is different from \cite{AMW,AW}, in which we remark again that the $\sigma$ is treated over $\mathbb{F}_p$.
 It seems that we cannot settle the case when $r>2$ and leave it open.

The linear complexity is an important cryptographic characteristic of sequences
and provides information on predictability and thus unsuitability for cryptography. Here we give a short introduction to the linear complexity of periodic sequences.
Let $\mathbb{F}$ be a field.  For a $T$-periodic
sequence $(s_n)$ over $\mathbb{F}$, recall that the
\emph{linear complexity} over $\mathbb{F}$, denoted by  $LC^{\mathbb{F}}((s_n))$, is the least order $L$ of a linear
recurrence relation over $\mathbb{F}$
$$
s_{n+L} = c_{L-1}s_{n+L-1} +\ldots +c_1s_{n+1}+ c_0s_n\quad
\mathrm{for}\,\ n \geq 0,
$$
which is satisfied by $(s_n)$ and where $c_0\neq 0, c_1, \ldots,
c_{L-1}\in \mathbb{F}$.
Let
$$
S(X)=s_0+s_1X+s_2X^2+\ldots+s_{T-1}X^{T-1}\in \mathbb{F}[X],
$$
which is called the \emph{generating polynomial} of $(s_n)$. Then the linear
complexity over $\mathbb{F}$ of $(s_n)$ can be computed as
\begin{equation}\label{licom}
  LC^{\mathbb{F}}((s_n)) =T-\deg\left(\mathrm{gcd}(X^T-1,
  ~S(X))\right),
\end{equation}
which is the degree of the \emph{characteristic polynomial}, $\frac{X^T-1}{\mathrm{gcd}(X^T-1,
  ~S(X))}$  , of the sequence.
See, e.g., \cite{CDR} for details.

For a sequence to be cryptographically strong, its linear complexity
should be high, but this complexity is not significantly reduced by changing a few
terms. This leads to the notion of the $k$-error linear complexity.
For integers $k\ge 0$, the \emph{$k$-error linear complexity} over $\mathbb{F}$ of $(s_n)$, denoted by $LC^{\mathbb{F}}_k((s_n))$, is the lowest linear complexity (over $\mathbb{F}$) that can be
obtained by changing at most $k$ terms of the sequence per period (see \cite{SM}, and see \cite{DXS} for the related sphere complexity that was defined even earlier).  Clearly, $LC^{\mathbb{F}}_0((s_n))=LC^{\mathbb{F}}((s_n))$, and
$$
T\ge LC^{\mathbb{F}}_0((s_n))\ge LC^{\mathbb{F}}_1((s_n))\ge \ldots \ge LC^{\mathbb{F}}_w((s_n))=0
$$
when $w$ equals the number of nonzero terms of $(s_n)$ per period, i.e., the weight of $(s_n)$.

\section{A lower bound on linear complexity}

In this section, we prove a lower bound on linear complexity of $\sigma$ in Eq.(\ref{chi}) for $q=p^r$ and $r\geq 2$.
Some results have been given in \cite{MW01,W04}. Our bound in Theorem \ref{LC-lowerbound} below improves that in \cite{MW01,W04} greatly.

Let $\mathrm{ord}_{m}(2)$ denote the \emph{order} of $2$ modulo $m$, i.e.,
 $\mathrm{ord}_{m}(2)$ is the least positive integer such that $2^{\mathrm{ord}_{m}(2)}\equiv 1 \pmod{m}$.

\begin{lemma}\label{lambda-pr}
Let  $\mathrm{ord}_{p}(2)=\lambda$ with $1<\lambda<p$. If $2^{p-1}\not\equiv 1 \pmod {p^2}$, then  $\mathrm{ord}_{p^r}(2)=\lambda {p^{r-1}}$ for $r\geq 2$.
\end{lemma}
Proof.  We suppose $\mathrm{ord}_{p^r}(2)=\mu$ . First, we write $ 2^{\lambda}=1+ap$ for some integer $a$ since $ 2^{\lambda}  \equiv 1\pmod {p}$.
Then we have
 $$
 2^{\lambda p^{r-1}}=(1+ap)^{p^{r-1}}=1+ap^r+\ldots
 $$
 and hence  $2^{\lambda p^{r-1}} \equiv 1\pmod {p^r}$. This implies that $\mu$ is a divisor of $\lambda p^{r-1}$. On the other hand, since  $2^{p-1}\not\equiv 1 \pmod {p^2}$, we see that $p\nmid a$.

Second, the assumption of $2^{p-1}\not\equiv 1 \pmod {p^2}$ implies that
$\mu\geq p$. So we can write $\mu=\lambda p^{w}$ for some positive integer $w\leq r-1$. Suppose $w<r-1$. Then we derive
 $$
 2^{\lambda p^{w}}=(1+ap)^{p^{w}}=1+ap^{w+1}+\ldots,
 $$
which contradicts to $2^{\mu}=2^{\lambda p^{w}}\equiv 1 \pmod {p^r}$. Then $w=r-1$ and hence $\mu=\lambda p^{r-1}$.
So we finish the proof. \qed\\

\begin{theorem}\label{LC-lowerbound}
Let $\sigma$ be the binary sequence of period $q$ defined in Eq.(\ref{chi}) with  $q=p^{r}$ for $r\geq 2$.
If $2^{p-1} \not \equiv 1 \pmod{p^2}$, then
the linear complexity  of $\sigma$  satisfies
$$
LC^{\mathbb{F}_2}(\sigma)\geq \lambda p^{r-1},
$$
where $1<\lambda< p$ is the order of $2$ modulo $p$.
\end{theorem}
Proof. From Eq.(\ref{chi}), it is easy to see that the least period of $\sigma$ is $q=p^r$, since there are $(q-1)/2$ many 1's in the first $q$ terms of the sequence.

 Let $\Phi^{(r)}(X)=1+X^{p^{r-1}}+X^{2p^{r-1}}+\ldots+X^{(p-1)p^{r-1}}\in  \mathbb{F}_2[X]$.
We see that $X^{p^{r}}-1=(X^{p^{r-1}}-1)\cdot \Phi^{(r)}(X)$ and $\Phi^{(r)}(X)$ has exactly $p^r-p^{r-1}$ many roots, which are  $p^r$-th primitive elements in $\overline{\mathbb{F}}_2$.
Then by Lemma \ref{lambda-pr}, $\Phi^{(r)}(X)$ can be written as
the product of $(p-1)/\lambda$ many irreduciable polynomials of degree $\lambda p^{r-1}$:
$$
\Phi^{(r)}(X)=\varphi^{(r)}_1(X)\varphi^{(r)}_2(X)\cdots \varphi^{(r)}_{(p-1)/\lambda}(X).
$$
We show below that there exists $i_0: 1\leq i_0\leq (p-1)/\lambda$ such that $\varphi^{(r)}_{i_0}(X)\nmid S(X)$, where $S(X)$ is the generating polynomial of $\sigma$.

Now  if we suppose $\Phi^{(r)}(X)|S(X)$ and write $S(X)=H(X)\cdot \Phi^{(r)}(X)$ for some polynomial $H(X)=h_0+h_1X+\ldots+h_{p^{r-1}-1}X^{p^{r-1}-1}\in \mathbb{F}_2[X]$ of degree $<p^{r-1}$.
Then we derive for $0\leq i<p^{r-1}$
$$
\sigma_{i}=\sigma_{i+p^{r-1}}=\ldots = \sigma_{i+(p-1)p^{r-1}}=
\left\{
\begin{array}{ll}
0,& \mathrm{if}\quad h_i=0,\\
1,& \mathrm{if}\quad h_i=1,
\end{array}
\right.
$$
from which we get $\sigma_n=\sigma_{n+p^{r-1}}$ for any integer $n$ and hence $p^{r-1}$ is the period of $\sigma$, a contradiction.
Hence $\Phi^{(r)}(X)\nmid S(X)$ and there exists at least one  $\varphi^{(r)}_{i_0}(X)$ such that $\varphi^{(r)}_{i_0}(X)\nmid S(X)$.
Then from the notion of the characteristic polynomial of $\sigma$ or Eq.(\ref{licom}), we have
$$
LC^{\mathbb{F}_2}(\sigma)\geq \deg(\varphi^{(r)}_{i_0}(X))=\lambda p^{r-1}.
$$
 we finish the proof. \qed\\

The bound is much better than that of \cite[Thms.1 and 2]{MW01} and \cite{W04}.  We note that, Theorem \ref{LC-lowerbound} 
is indeed a general result for any $p^r$-periodic binary sequences over $\mathbb{F}_2$ and it 
covers almost all primes.
 As far as we know, the primes that satisfy $2^{p-1}   \equiv 1 \pmod{p^2}$ are very rare.
It was shown  that there are only two such primes\footnote{A prime $p$ satisfying $2^{p-1}   \equiv 1 \pmod{p^2}$ is called a Wieferich prime.}, 1093 and 3511,
up to $6 \times 10^{17}$ \cite{AS}.

We remark again, in \cite[Prop.2]{BEP} for any sequences over $\mathbb{F}_{p^m}$ with least period $p^r$, the linear complexity is at least $p^{r-1}+1$. 
Theorem \ref{LC-lowerbound} is a very similar statement to \cite{BEP} for binary sequences.

\section{$k$-Error linear complexity}

In this section we consider the $k$-error linear complexity of $\sigma$ in Eq.(\ref{chi}) for $q=p^2$.

The way in the proof of Theorem \ref{LC-lowerbound} can help us to give a lower bound on the $k$-error linear complexity.
Below we choose $\{1,\gamma\}$ as a basis of $\mathbb{F}_{p^2}$ over $\mathbb{F}_p$ and write $\xi_n\in \mathbb{F}_{p^2}$ as $n_1+n_2\gamma$ for $n=n_1+n_2p$, where $0\leq n_1,n_2<p$.
We first prove two lemmas.

\begin{lemma}\label{tt}
Let $T_i=\{i+j\gamma : 0\leq j<p\}\subseteq \mathbb{F}_{p^2}$ for $0\leq i<p$. Then we have
$$
i\cdot T_1:=\{i(1+j\gamma) : 0\leq j<p\}=T_i, ~~~ i=1,2,\ldots,p-1.
$$
\end{lemma}
Proof. For each $1\leq i<p$, when $j$ runs through the set $\{0,1,\ldots,p-1\}$, so does $[ij]$, where  $[ij]$ is  $ij$ modulo $p$.
So $i(1+j\gamma)=i+[ij]\gamma\in T_i$ and $i(1+j_1\gamma)\neq i(1+j_2\gamma)$ if $0\leq j_1\neq j_2<p$. We finish the proof. \qed

\begin{lemma}\label{vector}
Let the vector $\overrightarrow{v_i}:=(\sigma_i, \sigma_{i+p}, \ldots, \sigma_{i+(p-1)p})$ for $0\leq i<p$, where the $\sigma_n$ is defined by Eq.(\ref{chi}) with $q=p^2$.
Let $\xi_{n_1+n_2p}=n_1+n_2\gamma \in \mathbb{F}_{p^2}$ in Eq.(\ref{chi}) be defined by using a basis $\{1,\gamma\}$  over $\mathbb{F}_p$ for $0\leq n_1,n_2<p$.
Let $wt(\overrightarrow{v_i})$, i.e. the weight of $\overrightarrow{v_i}$, denote the number of 1's in $\overrightarrow{v}_i$.

(1). If $\chi(\gamma)=-1$, we have $wt(\overrightarrow{v_0})=p-1$ and  $wt(\overrightarrow{v_i})=(p-1)/2$ for $1\leq i<p$;

(2). If $\chi(\gamma)=1$, we have $wt(\overrightarrow{v_0})=0$ and  $wt(\overrightarrow{v_i})=(p+1)/2$ for $1\leq i<p$.
\end{lemma}
Proof.
We first show $\chi(i)=1$ for any $i\in \{1,2,\ldots,p-1\}\subseteq \mathbb{F}_{p^2}$. Since $\chi$ is the quadratic character of $\mathbb{F}_{p^2}$, we can write $\chi=\eta^{(p^2-1)/2}$,
 where $\eta$ is a character of order $p^2-1$  of $\mathbb{F}_{p^2}$. Then from $i^{(p^2-1)/2}=(i^{p-1})^{(p+1)/2}=1+ap$ for some integer $a$,
we have
$$
\chi(i)=\eta(i^{(p^2-1)/2})=\eta(1)=1.
$$

Now from Eq.(\ref{chi}), we see that $\chi(\xi_n)=(-1)^{\sigma_n}$ for any $0\neq \xi_n\in \mathbb{F}_{p^2}$. Hence for $1\leq i\neq j<p$,
$wt(\overrightarrow{v_i})=wt(\overrightarrow{v_j})$ by Lemma \ref{tt}. While from $\chi(\xi_{jp})=\chi(j\gamma)=\chi(j)\chi(\gamma)=\chi(\gamma)$ for $1\leq j<p$,
we derive $wt(\overrightarrow{v_0})=p-1$ if $\chi(\gamma)=-1$ and otherwise  $wt(\overrightarrow{v_0})=0$.

Finally, since there are $(p^2-1)/2$ many $\xi_n\in \mathbb{F}_{p^2}$ such that $\chi(\xi_n)=1$, we have
$\sum\limits_{0\leq i<p}wt(\overrightarrow{v_i})=(p^2-1)/2$
and hence  $wt(\overrightarrow{v_i})=(p-1)/2$ for $1\leq i<p$ if $\chi(\gamma)=-1$, and otherwise
$wt(\overrightarrow{v_i})=(p+1)/2$ for $1\leq i<p$.
\qed\\

For $\sigma$ defined in Eq.(\ref{chi}) with $q=p^2$, write
$$
V_i(X)=\sum\limits_{j=0}^{p-1}\sigma_{i+jp} X^{i+jp}
$$
for $0\leq i<p$. Then clearly the generating polynomial $S(X)$ of $\sigma$ is $S(X)=\sum\limits_{i=0}^{p-1}V_i(X)$.

\begin{theorem}\label{lowerbound}
Let $\sigma$ be the binary sequence of period $q$ defined in Eq.(\ref{chi}) with  $q=p^{2}$.
Let $\xi_{n_1+n_2p}=n_1+n_2\gamma \in \mathbb{F}_{p^2}$ in Eq.(\ref{chi}) be defined by using a basis $\{1,\gamma\}$  over $\mathbb{F}_p$ for $0\leq n_1,n_2<p$.
If $2^{p-1} \not \equiv 1 \pmod{p^2}$, then
the $k$-error linear complexity  of $\sigma$  satisfies
$$
LC^{\mathbb{F}_2}_k(\sigma)\geq \lambda p,
$$
where $1<\lambda< p$ is the order of $2$ modulo $p$ and
$$
  0\leq k<
\left\{
\begin{array}{ll}
(p-1)^2/2,& \mathrm{if}\quad \chi(\gamma)=1,\\
1+(p-1)^2/2,& \mathrm{if}\quad \chi(\gamma)=-1.
\end{array}
\right.
$$
\end{theorem}
Proof.
By Lemma \ref{lambda-pr}, $\Phi^{(2)}(X)=1+X^p+X^{2p}+\ldots+X^{(p-1)p}\in  \mathbb{F}_2[X]$ is
the product of $(p-1)/\lambda$ many irreduciable polynomials of degree $\lambda p$:
$$
\Phi^{(2)}(X)=\varphi^{(2)}_1(X)\varphi^{(2)}_2(X)\cdots \varphi^{(2)}_{(p-1)/\lambda}(X).
$$

Let $S_k(X)$ be a polynomial of degree smaller than $p^2$ over $\mathbb{F}_2$. We restrict that $S_k(X)$ has $k$ many different terms from $S(X)$, the generating polynomial of $\sigma$,
that is, if we write
$$
S_k(X)=S(X)+E(X)\in \mathbb{F}_2[X],
$$
then $E(X)$,  a polynomial of degree smaller than $p^2$, has exactly $k$ many monomials.
We want to find an $E(X)$ with smallest $k$ such that $\Phi^{(2)}(X)|S_k(X)$.

We suppose $S_k(X)=S(X)+E(X)=h(X)\Phi^{(2)}(X)$ for some $h(X)=h_0+h_1X+\ldots+h_{p-1}X^{p-1}$ of degree smaller than $p$ over $\mathbb{F}_2$.
We derive
\begin{equation}\label{EEE}
E(X)=\sum\limits_{0\leq i<p} \left(h_iX^{i}\Phi^{(2)}(X)-V_i(X)\right).
\end{equation}
We see that $E(X)$ contains a summation $V_i(X)$ if $h_i=0$ and $X^{i}\Phi^{(2)}(X)-V_i(X)$ otherwise for $0\leq i<p$ from Eq.(\ref{EEE}) above.

If $\chi(\gamma)=1$, we have $wt(\overrightarrow{v_0})=0$ and  $wt(\overrightarrow{v_i})=(p+1)/2$ for $1\leq i<p$ by Lemma \ref{vector},
that is, $V_0(X)=0$ and $V_i(X)$ has $(p+1)/2$ many terms for $1\leq i<p$.
So we can verify that the $E(X)$ below
$$
E(X)=\sum\limits_{1\leq i<p}\left(X^{i}\Phi^{(2)}(X)-V_i(X)\right)
$$
is with smallest $k=(p-1)^2/2$ terms (such that $\Phi^{(2)}(X)|S_k(X)$).
The argument tells us that if $k<(p-1)^2/2$, no $E(X)$ with $k$ terms can guarantee $\Phi^{(2)}(X)|S_k(X)$.
Then for such $k$, at least one of $\varphi^{(2)}_1(X), \varphi^{(2)}_2(X), \cdots, \varphi^{(2)}_{(p-1)/\lambda}(X)$ is not a divisor of $S_k(X)$,
and hence $LC^{\mathbb{F}_2}_k(\sigma)\geq \lambda p$ by Eq.(\ref{licom}).

If $\chi(\gamma)=-1$, Lemma \ref{vector} helps us to verify that the $E(X)$ below
$$
E(X)=\Phi^{(2)}(X)-V_0(X)+\sum\limits_{1\leq i<p}V_i(X)=1+\sum\limits_{1\leq i<p}V_i(X)
$$
is with smallest $k=1+(p-1)^2/2$ terms such that $\Phi^{(2)}(X)|S_k(X)$. Then following the way above, we finish the proof. \qed\\

Now we consider the case when $2$ is primitive modulo $p^2$.
We need the following lemma.

\begin{lemma}\label{poly}
Let $V_i(X)=\sum\limits_{0\leq j<p}\sigma_{i+jp}X^{i+jp}\in \mathbb{F}_2[X]$  for $0\leq i<p$, where the $\sigma_n$ is defined by Eq.(\ref{chi}) with $q=p^2$.
Let $\xi_{n_1+n_2p}=n_1+n_2\gamma \in \mathbb{F}_{p^2}$ in Eq.(\ref{chi}) be defined by using a basis $\{1,\gamma\}$  over $\mathbb{F}_p$ for $0\leq n_1,n_2<p$.
We have
$$
V_0(X)\equiv 0 \pmod{X^p-1}, ~~~
V_i(X)\equiv wX^i \pmod{X^p-1},
$$
where
$$
 w=
\left\{
\begin{array}{ll}
(p+1)/2,& \mathrm{if}\quad \chi(\gamma)=1,\\
(p-1)/2,& \mathrm{if}\quad \chi(\gamma)=-1.
\end{array}
\right.
$$
\end{lemma}
Proof. It is clear from Lemma \ref{vector}. \qed

\begin{theorem}\label{exactvalue=1}
Let $\sigma$ be the binary sequence of period $q$ defined in Eq.(\ref{chi}) with  $q=p^{2}$.
Let $\xi_{n_1+n_2p}=n_1+n_2\gamma \in \mathbb{F}_{p^2}$ in Eq.(\ref{chi}) be defined by using a basis $\{1,\gamma\}$  over $\mathbb{F}_p$ for $0\leq n_1,n_2<p$.
If $\chi(\gamma)=1$ and $2$ is primitive modulo $p^2$, then
the $k$-error linear complexity  of $\sigma$  satisfies
 \[
 LC^{\mathbb{F}_2}_k(\sigma)=\left\{
\begin{array}{cl}
p^2-1, & \mathrm{if}\,\ k=0, \\
p^2-p+1, & \mathrm{if}\,\ 1\le k<p-1, \\
p^2-p, & \mathrm{if}\,\ p-1\leq k<(p-1)^2/2,\\
p-1, & \mathrm{if}\,\ k=(p-1)^2/2,\\
0, & \mathrm{if}\,\  k\geq (p^2-1)/2,
\end{array}
\right.\\
\]
if $p\equiv 5 \pmod 8$, and
\[
 LC^{\mathbb{F}_2}_k(\sigma)=\left\{
\begin{array}{cl}
p^2-p, & \mathrm{if}\,\  0\leq k <(p-1)^2/2, \\
p-1, & \mathrm{if}\,\ k=(p-1)^2/2,\\
0, & \mathrm{if}\,\ k\geq (p^2-1)/2,
\end{array}
\right.\\
\]
if $p\equiv 3 \pmod 8$.
\end{theorem}
Proof. As before let $S(X)$ be the generating polynomial of $\sigma$.
We mention that
both $\Phi^{(2)}(X)=1+X^p+\ldots+X^{(p-1)p}$ and $1+X+\ldots+X^{p-1}$ are irreduciable, since $2$ is primitive modulo $p^2$.

From the proof of Theorem \ref{lowerbound}, we see that $\Phi^{(2)}(X)\nmid S(X)$ and for any $E(X)$ with $k<(p-1)^2/2$ many terms, $\Phi^{(2)}(X)\nmid S_k(X)$ for
$S_k(X)=S(X)+E(X)$.
We now consider $S(X)$ modulo $(X^p-1$). By Lemma \ref{poly}, we have
\begin{equation}\label{SmodX}
\begin{array}{rl}
S(X) \pmod {X^p-1} & \equiv \frac{p+1}{2}(X+X^2+\ldots+X^{p-1})\\
                   & \equiv \left\{
\begin{array}{ll}
X+X^2+\ldots+X^{p-1}, & \mathrm{if} ~ p\equiv 5 \pmod 8,\\
0, &  \mathrm{if} ~  p\equiv 3 \pmod 8.
\end{array}
\right.
\end{array}
\end{equation}
We consider the case when $p\equiv 5 \pmod 8$. From Eq.(\ref{SmodX}) we derive

(i).  $S(1)=0$ and $\gcd(S(X), (X^{p^2}-1)/(X-1))=1$, then we have $LC^{\mathbb{F}_2}_0(\sigma)=p^2-1$.

(ii). $S(X)+1 \equiv 1+X+\ldots+X^{p-1} \pmod {X^p-1}$, which indicates that $LC^{\mathbb{F}_2}_1(\sigma)=p^2-p+1$.

(iii). $S(X)+X+\ldots+X^{p-1} \equiv 0 \pmod {X^p-1}$, which indicates that $LC^{\mathbb{F}_2}_{p-1}(\sigma)=p^2-p$.

Putting everything together, we prove the first statement. For the  case when $p\equiv 3 \pmod 8$, we get easily that
 $LC^{\mathbb{F}_2}_0(\sigma)=p^2-p$ and then from arguments above,  we prove the second statement.
 \qed\\

Similarly, we have following theorem for  $\chi(\gamma)=-1$.
\begin{theorem}\label{exactvalue=-1}
Let $\sigma$ be the binary sequence of period $q$ defined in Eq.(\ref{chi}) with  $q=p^{2}$.
Let $\xi_{n_1+n_2p}=n_1+n_2\gamma \in \mathbb{F}_{p^2}$ in Eq.(\ref{chi}) be defined by using a basis $\{1,\gamma\}$  over $\mathbb{F}_p$ for $0\leq n_1,n_2<p$.
If $\chi(\gamma)=-1$ and $2$ is primitive modulo $p^2$, then
the $k$-error linear complexity  of $\sigma$  satisfies
\[
 LC^{\mathbb{F}_2}_k(\sigma)=\left\{
\begin{array}{cl}
p^2-p, & \mathrm{if}\,\  0\leq k <1+(p-1)^2/2, \\
p, & \mathrm{if}\,\ k=1+(p-1)^2/2,\\
0, & \mathrm{if}\,\ k\geq (p^2-1)/2,
\end{array}
\right.\\
\]
if $p\equiv 5 \pmod 8$, and
\[
 LC^{\mathbb{F}_2}_k(\sigma)=\left\{
\begin{array}{cl}
p^2-1, & \mathrm{if}\,\ k=0, \\
p^2-p+1, & \mathrm{if}\,\ 1\le k<p-1, \\
p^2-p, & \mathrm{if}\,\ p-1\leq k<1+(p-1)^2/2,\\
p, & \mathrm{if}\,\ k=1+(p-1)^2/2,\\
0, & \mathrm{if}\,\  k\geq (p^2-1)/2,
\end{array}
\right.\\
\]
if $p\equiv 3 \pmod 8$.
\end{theorem}

Theorems \ref{exactvalue=1} and \ref{exactvalue=-1} indicate that  the sequence we considered has good stability, or in other words, its linear complexity is not significantly
 decreased by changing only a few (but not many) terms.

\section{Final remarks}

We study the linear complexity of binary sequences defined by using the
 quadratic character of the finite field $\mathbb{F}_{p^r}$ with $r\geq 2$ and its $k$-error linear complexity for $r=2$. Such sequences are an extension of Legendre sequences.
It is interesting to consider the $k$-error linear complexity for $r>2$.

From the construction, we find by Lemma \ref{vector} that $\sigma_1=\sigma_2=\cdots=\sigma_{p-1}$ and $\sigma_{p}=\sigma_{2p}=\cdots=\sigma_{(p-1)p}$.
This sacrifices  some pseudorandomness of the sequence. So we can modify the construction as follows
$$
(-1)^{\sigma_n}=\left\{
\begin{array}{ll}
1,& \mathrm{if}\quad n= 0,\\
\left(\frac{j}{p}\right), & \mathrm{if}\quad n=jp \quad \mathrm{for}\quad 1\leq j<p, \\
\left(\frac{i}{p}\right)\chi(\xi_n), &\mathrm{if}\quad n=i+jp \quad \mathrm{for}\quad 1\leq i<p, 0\leq j<p.
\end{array}
\right.
$$
Then the way in this work can be used to consider the linear complexity and $k$-error linear complexity.

Finally we remark that, there is another way to order the elements in $\mathbb{F}_q$. Write $\mathbb{F}_q=\{0,1,\alpha,\alpha^{2},\ldots,\alpha^{q-2}\}$, where
 $\alpha$ is  a primitive element of $\mathbb{F}_q$.
The sequence $\rho=(\rho_0,\rho_1,\ldots)$ is defined by
$$
\rho_n=\left\{
\begin{array}{ll}
0,& \mathrm{if}\quad n=(q-1)/2,\\
(1-\chi(\alpha^{n}-1))/2,&\mathrm{otherwise}.
\end{array}
\right.
$$
$\rho$ is referred to as a \emph{generalized Sidelnikov sequence}, see e.g. \cite{AW}, in which the $k$-error linear complexity (over $\mathbb{F}_p$) of $\rho$
was determined when $r=1$. So it is interesting to consider the $k$-error linear complexity (over $\mathbb{F}_2$) of $\rho$.

\section*{Acknowledgment}

The authors wish to thank Prof. Arne Winterhof for helpful suggestions and some corrections of the proof.


The work was partially supported by the National Natural Science
Foundation of China under grant No.~61772292, by the Projects of International Cooperation and Exchanges NSFC No. 6181101289, by the Provincial Natural Science
Foundation of Fujian under grant No.~2018J01425 and by the Program for Innovative Research Team in Science and Technology in Fujian Province University under grant No.~2018-49.


\begin{thebibliography}{99}

\bibitem{AMW} Aly, H.,  Meidl, W.,  Winterhof, A.: On the $k$-error linear complexity of
cyclotomic sequences. J. Math. Crypt. 1 (2007) 283-296.


\bibitem{AW} Aly, H., Winterhof, A.: On the $k$-error linear complexity over
of Legendre and Sidelnikov sequences.  Designs, Codes and
Cryptography 40 (2006) 369-374.



\bibitem{AS} Akbary, A., Siavashi, S.: The largest known Wieferich numbers. Integers 18-\#A3 (2018)  1-6.

\bibitem{BEP} Blackburn, S. R., Etzion, T., Paterson, K. G.:
Permutation polynomials, de Bruijn sequences, and linear complexity. J. Comb. Theory Ser. A 76(1) (1996) 55-82.


\bibitem{CDR}
 Cusick, T. W., Ding, C., Renvall, A.: Stream Ciphers and Number Theory. Gulf Professional Publishing, 2004.





\bibitem{DXS}
Ding, C., Xiao, G., Shan, W.: The stability theory of stream
ciphers. Lecture Notes in Computer Science, vol.561, Berlin: Springer-Verlag (1991).


\bibitem{DHS} Ding, C., Helleseth, T., Shan, W.: On the linear
complexity of Legendre sequences.  IEEE Transactions on Information
Theory 44(3) (1998)  1276-1278.


\bibitem{D95} Ding, C.: Binary cyclotomic generators.  Fast
Software Encrytion,Lecture Notes in Computer Science, vol.1008. Berlin: Springer-Verlag (1995)
20-60.



\bibitem{D98-2} Ding, C.: Pattern distributions of Legendre
sequences. IEEE Transactions on Information Theory 44(4) (1998)
1693-1698.

\bibitem{GMS}
Gyarmati, K., Mauduit, C., S\'{a}rk\"{o}zy, A.:
On finite pseudorandom binary lattices. Discrete Applied Mathematics 216 (2017)  589-597.

\bibitem{GSS09} Gyarmati, K., S\'{a}rk\"{o}zy, A., Stewart, C. L.:
On Legendre symbol lattices. Unif. Distrib. Theory 4 (2009) 81-95.

\bibitem{GSS13}
Gyarmati, K., S\'{a}rk\"{o}zy, A., Stewart, C. L.: On Legendre symbol lattices, II. Unif. Distrib. Theory 8 (2013) 47-65.



\bibitem{KS} Kim, J. H.,  Song, H. Y.: Trace representation of Legendre sequences. Des. Codes Cryptogr. 24  (2001)  343-348.


\bibitem{MW01} Meidl, W., Winterhof, A.: Lower bounds on the linear complexity of the
discrete logarithm in finite fields. IEEE Transactions on
Information Theory 47(7)  (2001) 2807-2811.


\bibitem{MW032} Meidl, W., Winterhof, A.:  On the autocorrelation of cyclotomic
generator. Lecture Notes in Computer Science, vol.
2948, Springer-Verlag Berlin Heidelberg (2003)  1-11.





\bibitem{SW} S\'{a}rk\"{o}zy, A.,  Winterhof, A.: Measures of pseudorandomness
for binary sequences constructed using finite fields. Discrete
Mathematics 309(6) (2009) 1327-1333.


\bibitem{SM} Stamp, M., Martin, C. F.: An algorithm for the $k$-error linear complexity of
 binary sequences with period $2^n$. IEEE Trans. Inform. Theory 39(4) (1993) 1398-1401.



\bibitem{W04}  Winterhof,  A.: A note on the linear complexity profile of the discrete logarithm in finite fields.
in: K. Feng, H. Niederreiter, C. Xing (Eds.), Coding, Cryptography
and Combinatorics, Progr. Comput. Sci. Appl. Logic 23, Basel:
Birkh\"auser (2004) 359-367.



\end{thebibliography}
\end{document}